# Superconductivity at 22 K in Co-doped $BaFe_2As_2$ Crystals

Athena S. Sefat, Rongying Jin, Michael A. McGuire, Brian C. Sales, David J. Singh, David Mandrus

*Materials Science & Technology Division, Oak Ridge National Laboratory, Oak Ridge, TN 37831, USA*

Here we report bulk superconductivity in $BaFe_{1.8}Co_{0.2}As_2$ single crystals below $T_c$ = 22 K, as demonstrated by resistivity, magnetic susceptibility, and specific heat data. Hall data indicate that the dominant carriers are electrons, as expected from simple chemical reasoning. This is the first example of superconductivity induced by electron doping in this family of materials. In contrast to the cuprates, the $BaFe_2As_2$ system appears to tolerate considerable disorder in the FeAs planes. First principles calculations for $BaFe_{1.8}Co_{0.2}As_2$ indicate the inter-band scattering due to Co is weak.

The recent discovery of high superconducting transition temperatures in FeAs based compounds has generated great interest in the scientific community. In the oxypnictide system *R*FeAsO (*R* = Pr, Sm, Nd, Gd), electron- or hole-doping has produced critical temperatures ($T_c$) as high as 56 K [1-14]. In the oxygen-free compounds of $AFe_2As_2$ ($A$ = Ca, Sr, Ba), hole-doping has reached $T_c$ values of 38 K [15, 16]. This is the first report of superconductivity in $BaFe_2As_2$, induced by electron-doping with cobalt on the Fe site. Since superconductivity in FeAs based compounds coincides with the disappearance of a spin-density-wave type magnetic transition, spin fluctuations of Fe are suggested to be important in developing the superconducting ground state [17]. It is found that by introducing holes in the $(FeAs)^-$ layers, by K-doping in $BaFe_2As_2$ for example, the structural and magnetic phase transition at approximately 140 K is suppressed and superconductivity is found [15, 18]. This scenario is very similar to the arsenide oxide superconductors [19, 20]. Chemically, cobalt has advantages as a dopant since carriers are added directly into the FeAs planes, and cobalt is much easier to handle than alkali metals. We have recently reported bulk superconductivity for ~ 4 - 12 % Co doping levels in LaFeAsO, giving onset transition temperature up to $T_c^{onset}$ = 14.3 K in $LaFe_{0.92}Co_{0.08}AsO$ [14]. Here we report superconductivity in 8.0(5) % cobalt doped $BaFe_2As_2$ single crystals, giving $T_c^{onset}$ = 22 K. The relatively high $T_c$ is exciting as it demonstrates that in-plane disorder is highly tolerated.

Large single crystals of $BaFe_2As_2$ and Co-doped $BaFe_2As_2$ were grown out of FeAs flux, allowing for intrinsic property investigation. FeAs flux is preferred over other metal solvents such as Sn, as flux impurities become incorporated in the crystals. The typical crystal sizes from both batches were ~ 8 x 5 x 0.2 $mm^3$. In the preparation of crystals, high purity elements (> 99.9 %) were used and the source of all elements was Alfa Aesar. First, CoAs and FeAs binaries were prepared by placing mixtures of As, Co powder and/or Fe pieces in a silica tube. These were reacted slowly by heating to 700 °C (dwell 6 hrs), then to 1065 °C (dwell 10 hrs). For $BaFe_2As_2$ crystals, a ratio of Ba:FeAs = 1:5 was heated for 8 hours at 1180°C under partial atm argon. For Co-doped $BaFe_2As_2$, a ratio of Ba:FeAs:CoAs = 1:4.45:0.55 was heated to 1180 °C, and held for 10 hours. In both cases the ampoules were cooled at the rate of 3 - 4 °C/hour, followed by decanting of FeAs flux at 1090 °C. The crystals were brittle, well-formed plates with the [001] direction perpendicular to the plane of the crystals.

Electron probe microanalysis of a cleaved surface of the single crystal was performed on a JEOL JSM-840 scanning electron microscope (SEM) using an accelerating voltage of 15 kV and a current of 20 nA with an EDAX brand energy-dispersive X-ray spectroscopy (EDS) device. EDS analyses on several crystals indicated that 8.0(5) % of the Fe is replaced by Co in $BaFe_2As_2$. Thus, the composition will be presented as $BaFe_{1.8}Co_{0.2}As_2$.

The phase purity of the crystals was determined using a Scintag XDS 2000 2θ-2θ diffractometer (Cu $K_\alpha$ radiation). The parent and Co-doped $BaFe_2As_2$ crystallize with the $ThCr_2Si_2$ [21, 22] structure, in tetragonal space group symmetry *I*4/*mmm* (No. 139; Z = 2) at room temperature. The crystal structure is made up of $Ba_{0.5}^{2+}(FeAs)^-$ for the parent, with partial and random substitution of Co on Fe sites for $BaFe_{1.8}Co_{0.2}As_2$. All observed reflections were indexed, limiting impurity concentrations to about 1% or less. Lattice constants were determined from LeBail refinements using the program FullProf [23]. The lattice parameters of $BaFe_2As_2$ are $a$ = 3.9635(5) Å and $c$ = 13.022(2) Å, consistent with a report on a polycrystalline sample [21]. The refined lattice constants of $BaFe_{1.8}Co_{0.2}As_2$ are $a$ = 3.9639(4) Å and $c$ = 12.980(1) Å. Cobalt doping results in a small decrease (0.26 %) in the length of the $c$-axis, while the value of $a$-axis is unchanged within experimental uncertainty. We have recently reported similar behavior in Co-doped LaFeAsO [14].

DC magnetization was measured as a function of temperature and field using a Quantum Design Magnetic Property Measurement System. Fig. 1a shows the measured magnetic susceptibility ($\chi$) in zero-field-cooled (*zfc*) form for $BaFe_2As_2$ in 1 kOe applied field along *c*- and *ab*-crystallographic directions. The susceptibility is anisotropic and decreases slightly with decreasing temperature down to ≈ 140 K, at which point it drops more abruptly, due to a magnetic and structural phase transition [21]. As can be seen in Fig. 1b, $\chi$ for $BaFe_{1.8}Co_{0.2}As_2$ varies smoothly between ~ 22 K and 300 K without any abrupt drop, indicating the disappearance of magnetic ordering in the doped system.

Shown in Fig. 1c is the $4\pi\chi_{eff}$ versus T measured along *ab*-plane at an applied field of 20 Oe for $BaFe_{1.8}Co_{0.2}As_2$, where $\chi_{eff}$ is the effective magnetic susceptibility after demagnetization effect correction. Since the magnetization varied linearly with fields below 100 Oe at 1.8 K, we assume that the system is fully shielded; therefore, the effective magnetic susceptibility for *zfc* data is $\chi_{eff} = -1/(4\pi)$. The demagnetization factor N was then obtained via $N = (\chi^{-1} - \chi_{eff}^{-1})$; N = 11.3 for H // *c* and 5.7 for H //*ab*. The $\chi_{eff}$ at different temperatures was calculated using N and measured $\chi$ via $\chi_{eff} = \chi/(1-N\chi)$. Note that the *zfc* $4\pi\chi_{eff}$ data decreases rapidly for temperatures below ~ 22 K and quickly saturates to 1. This indicates that the system is indeed fully shielded and the substitution of cobalt on Fe-site is more homogenous than the use of other dopants such as fluorine in $LaFeAsO_{1-x}F_x$ [2] or potassium in $Ba_{1-x}K_xFe_2As_2$ [15] for which the transitions are broadened. However, the *fc* $4\pi\chi_{eff}$ saturates at much smaller value (< 0.01), probably reflecting strong pinning due to the Co dopant. This is further confirmed by M versus H at 1.8 K (Fig. 1c). Magnetization versus magnetic field deviates from linear behavior above 200 Oe, suggesting that the lower critical field $H_{c1}$ is approximately 200 Oe. However, M(H) does not reach zero in applied fields up to 7 T, indicating that the upper critical field $H_{c2}$ is much higher than 7 T. Nevertheless, one may estimate the critical current by measuring the magnetization hysteresis loop as shown in Fig. 1d. Using the Bean critical state model [24], we obtain a $J_c$ value along the *ab*-plane of ≈ 1.8 x $10^4$ A $cm^{-2}$. This large $J_c$ results from strong pinning effects. Although the magnetization hysteresis loop is also measured along the *c*-axis, it is non-trivial to calculate $J_c$, as H is not along symmetry axis [25]. In view of the magnetization loops, we note that in addition to the peak penetration field, $H_p$, there is another peak that occurs at $H_x$, when the magnetic field increases along both the *c*- and *ab*-crystallographic directions. At present it is unclear whether this is due to the fish tail effect as observed in high-$T_c$ cuprates [26], and more measurements are underway.

Temperature dependent electrical resistivity was performed on a Quantum Design Physical Property Measurement System (PPMS), measured in the *ab*-plane. For $BaFe_2As_2$ (inset of Fig. 2a), the room temperature $\rho_{300\ K}$ = 5.9 mΩ cm; this value is comparable to that reported [18]. The resistivity drops below ~ 150 K with decreasing temperature, reaching $\rho_{2K}$ = 5.0 mΩ cm. While both $BaFe_2As_2$ and $BaFe_{1.8}Co_{0.2}As_2$ show metallic behavior, the resistivity for $BaFe_{1.8}Co_{0.2}As_2$ is much smaller than for the parent compound. For $BaFe_{1.8}Co_{0.2}As_2$ as the temperature is lowered down to 22 K, an abrupt drop in $\rho_{ab}$ is observed (Fig. 2a). The onset transition temperature for 90 % fixed percentage of the normal-state value is $T_c^{onset}$ = 22 K at 0 T. The transition width in 0 T is $\Delta T_c = T_c$ (90%) - $T_c$ (10%) = 0.6 K. The $\Delta T_c$ value is much smaller that that reported for $LaFe_{0.92}Co_{0.08}AsO$ ($\Delta T_c$ = 2.3 K) [14], and for $LaFeAsO_{0.89}F_{0.11}$ ($\Delta T_c$ = 4.5 K) [2]. The resistive transition shifts to lower temperatures by applying an 8 T magnetic field, and transition width becomes wider (1.3 K), a characteristic of type-II superconductivity.

For the Hall-effect measurements, a rectangular-shaped single-crystalline platelet of $BaFe_{1.8}Co_{0.2}As_2$ was selected with a thickness of 0.05 mm. Four electrical contacts were made using silver epoxy. The Hall voltage was calculated from the anti-symmetric part of the transverse voltage (perpendicular to the applied current) under magnetic-field reversal at both fixed field and temperature. Shown in Fig. 2b is the temperature dependence of the Hall coefficient $R_H$ taken at 8 T. Note that $R_H$ is negative between 30 and 300 K, initially becoming more negative with decrease of temperature. At lower temperatures, it becomes less negative after reaching a minimum near 100 K. As demonstrated in the inset of Fig. 2b, $\rho_H$ varies nearly linearly with H at various fixed temperatures (30, 100, 200, and 300 K), indicating no significant magnetic contributions to the measured Hall voltages. The negative $R_H$ indicates that electrons are the dominant charge carriers. The Hall coefficient of $BaFe_{1.8}Co_{0.2}As_2$ is smaller in absolute value than reported values for the superconducting oxypnictides [2]. The carrier concentration *n* may be inferred via $n = 1/(qR_H)$; *q* is the carrier charge. This gives ~ 6 x $10^{21}$ $cm^{-3}$ at 300 K. Interpretation of $R_H$ is complicated by the multi-band nature of the system and the presence of both electron and hole bands at the Fermi level [27, 28]. In a simple two band model, *n* derived from $R_H$ as outlined above represents an upper bound on the actual concentration of electrons. The temperature dependence of *n* may be more closely related to changes in relaxation rates associated with different electron and hole pockets than to the changes in carrier concentration.

Specific heat data, $C_p(T)$, were also obtained using the PPMS via the relaxation method. Fig. 2c gives the temperature dependence of specific heat for $BaFe_{1.8}Co_{0.2}As_2$. We clearly observe a specific heat discontinuity below $T_c$, peaking at 20 K. Below ~ 5 K, C/T has a linear $T^2$ dependence (inset of Fig. 2c). In the temperature range of 1.9 K to ~ 5 K, the fitted Sommerfeld coefficient, $\gamma$, is 0.47(1) mJ/($K^2$mol atom). This small $\gamma$ value may not be intrinsic to the superconducting phase, but could be due to a residue of metallic FeAs flux sticking to the crystal.

We performed density functional calculations for Co-doped $BaFe_2As_2$ within the Local Density Approximation (LDA), with methodology as described in Ref. [28] for the $BaFe_2As_2$ parent. These were done both for a virtual crystal with Fe replaced by a virtual atom $Z$ = 26.1 to simulate 10% Co doping, and in supercells with $Ba_2Fe_3CoAs_4$ and $Ba_4Fe_7CoAs_8$ compositions. In both cases the lattice parameters were held fixed and all internal coordinates were relaxed. The projections of the *d*-electron DOS for the Fe and Co in a $Ba_4Fe_7CoAs_8$ supercell is shown in Fig. 3a. Note the downward shift of the states that comprise the hole sections on the Co sites. Co is more strongly hybridized with As than Fe. The calculated Fermi surface of the virtual crystal without magnetism is shown in Fig. 3b. As may be seen in comparison with Ref. [28], the effect of Co doping on the structure of the Fermi surface at the virtual crystal level is rather similar to the effect of electron-doping on the Ba site. Turning to the supercell calculations, we find that Co is expected to strongly scatter carriers near the Fermi energy $E_F$. In particular, comparing the projections of the density of states (DOS) onto the Co and Fe sites we find that in the vicinity of the $E_F$ the Co contribution is less that 50% of that of an Fe atom. This depletion in Co contribution is even stronger below $E_F$ but the Co contribution becomes essentially the same as the Fe contribution starting from 0.2 eV above $E_F$. This is associated with a large pseudogap in the *d* electron DOS on the Co site associated with As-Co hybridization. These Fe-based materials are low carrier density metals, with heavy hole Fermi surfaces at the zone center and lighter electron sections around the zone corner [28]. Our supercell results show that Co much more strongly affects the states forming the hole sections, which have different orbital character than the electron sections. The implications are first of all that the scattering will be stronger on the hole sections, though for low **q** strong scattering is inevitable on both sections, and more significantly that inter-band scattering may be relatively weak. This is in analogy with the effect of defects on $MgB_2$, which shows two band superconductivity [29].

In summary we have found superconductivity in Co-doped $BaFe_2As_2$ single crystals with onset superconducting transition temperature of 22 K. The parent $BaFe_2As_2$ orders antiferromagnetically ($T_N \approx 140$ K) [21]. For 8.0(5) % cobalt doping, magnetic order is destroyed and superconductivity is induced. The bulk nature of the superconductivity and the high quality of the samples are confirmed by the anomaly in specific heat data, and narrow transition widths in both resistivity and magnetic susceptibility measurements. Hall data indicate that doping with Co produces an electron doped superconductor. These experiments and in particular the robustness of superconductivity against Co doping, which must introduce strong intra-band scattering, shed significant light on the nature of superconductivity in these Fe-based superconductors. In particular, intra-band scattering will average the order parameter on a given Fermi surface. This will apply also for nested cylinders with similar orbital character and small **q** spacing. Thus order parameters with nodes or sign changes (such as an axial *p*-wave state, which does not have nodes) will be destroyed, and only *s* like states on a given set of Fermi surfaces (holes or electrons) will survive. Robustness of superconductivity in the presence of strong inter-band scattering would seem to lead to the same conclusion, i.e. an *s* symmetry superconducting state. However, while superconductivity remarkably survives Co doping both in this system and as reported by us earlier in the oxy-arsenides [14], the critical temperature in the oxy-arsenide is rather strongly reduced in Co-doped samples as compared to F-doped materials. In view of this and considering that the band structure results suggest that inter-band scattering due to Co may be relatively weak (note that these are high $T_c$, low Fermi velocity, short coherence length materials), we do not exclude generalized *s*-state, e.g. where there is a phase shift or sign change between the order parameters on the hole and electron surfaces. One such scenario involving spin-fluctuations was discussed by Mazin and co-workers [30].

## Acknowledgement

We are grateful for helpful discussions with I. I. Mazin. The research at ORNL was sponsored by the Division of Materials Sciences and Engineering, Office of Basic Energy Sciences, U.S. Department of Energy. Part of this research was performed by Eugene P. Wigner Fellows at ORNL, managed by UT-Battelle, LLC, for the U.S. DOE under Contract DE-AC05-00OR22725.

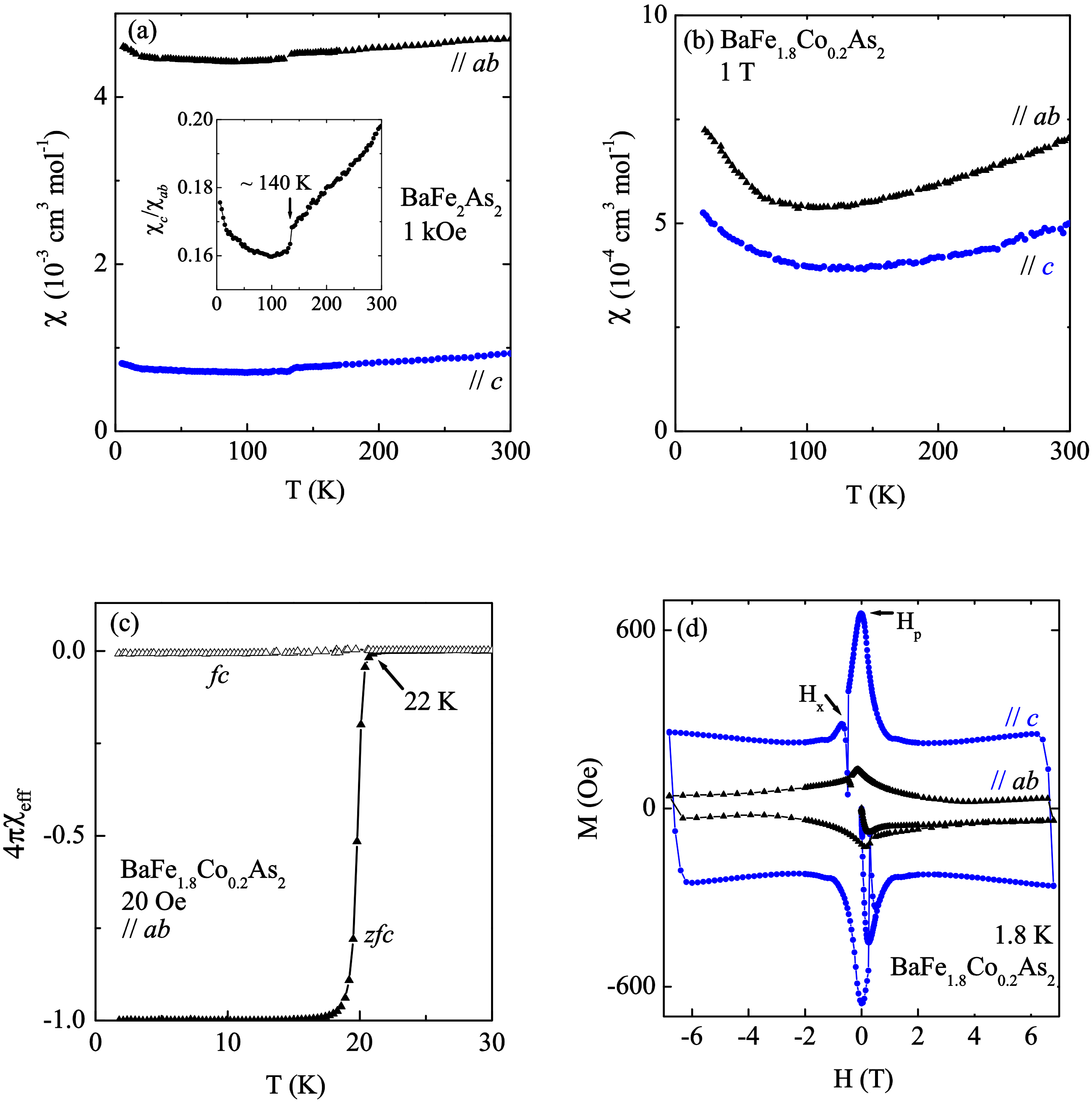

Fig. 1. (Color online) Temperature dependence of molar susceptibility in zero-field cooled (*zfc*) forms along the two crystallographic directions for (a) $BaFe_2As_2$ in 1 kOe and (b) $BaFe_{1.8}Co_{0.2}As_2$ in 1 T. (c) The temperature dependence of susceptibility in *zfc* and field-cooled (*fc*) forms below 30 K, for $BaFe_{1.8}Co_{0.2}As_2$ in 20 Oe along *ab*-plane, assuming $\chi$ value of the perfect diamagnetism. (d) Field dependence of magnetization for $BaFe_{1.8}Co_{0.2}As_2$ along the two crystallographic directions at 1.8 K; there are two peaks noted at $H_p$ and $H_x$. The demagnetization shape factor is accounted for in the $BaFe_{1.8}Co_{0.2}As_2$ superconductivity data (see text).

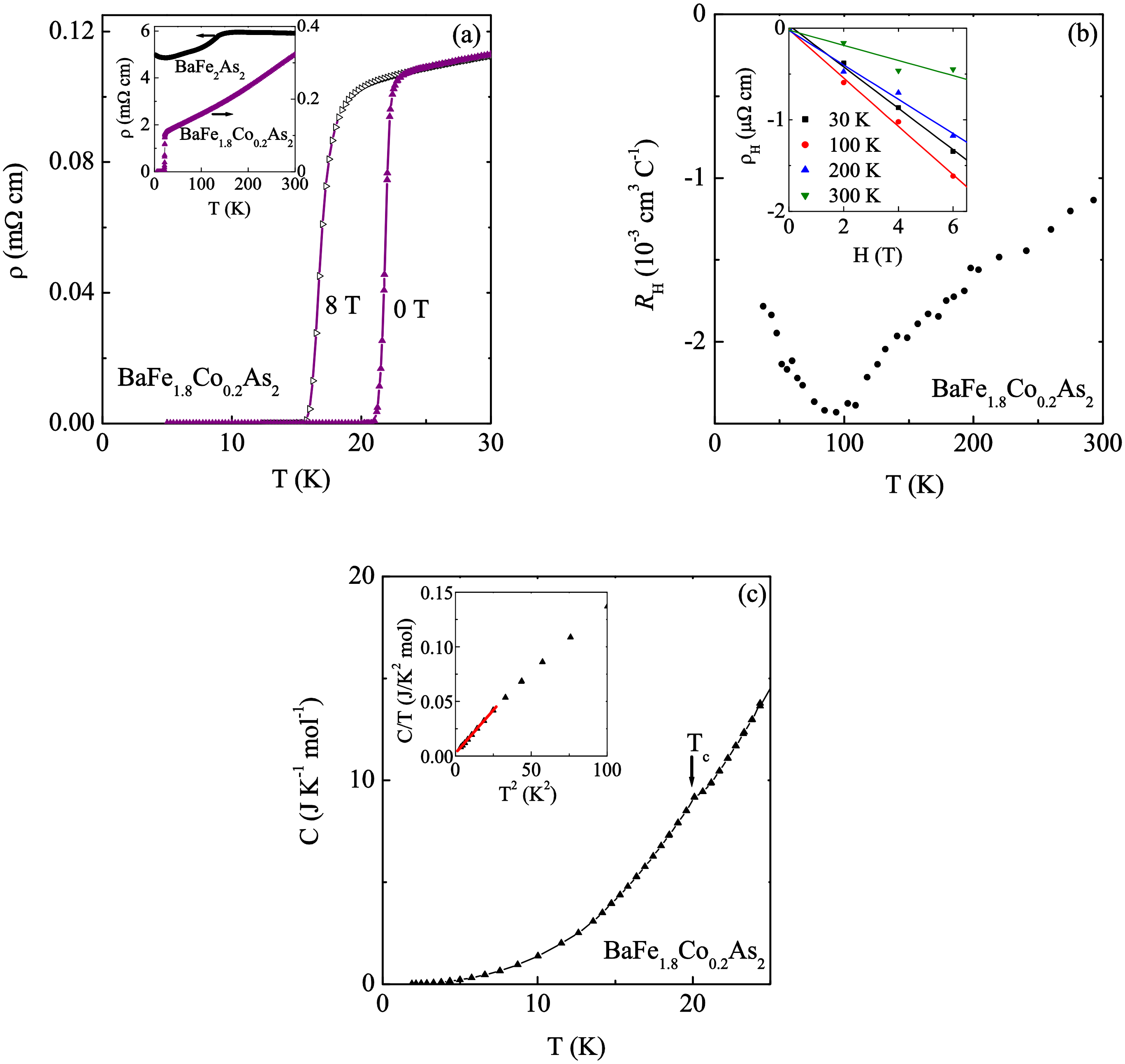

Fig. 2. (Color online) (a) Temperature dependence of resistivity for $BaFe_{1.8}Co_{0.2}As_2$, measured below 30 K, at 0 T and 8 T applied magnetic fields. Inset is the temperature dependence of $BaFe_2As_2$ and $BaFe_{1.8}Co_{0.2}As_2$ up to room temperature at 0 T. (b) Hall coefficient $R_H$ versus temperature for $BaFe_{1.8}Co_{0.2}As_2$. Inset is the field dependence of resistivity at 30, 100, 200, and 300 K. (c) Temperature dependence of specific heat for $BaFe_{1.8}Co_{0.2}As_2$. Inset is the C/T vs $T^2$ and linear fit below ~5 K.

(a)

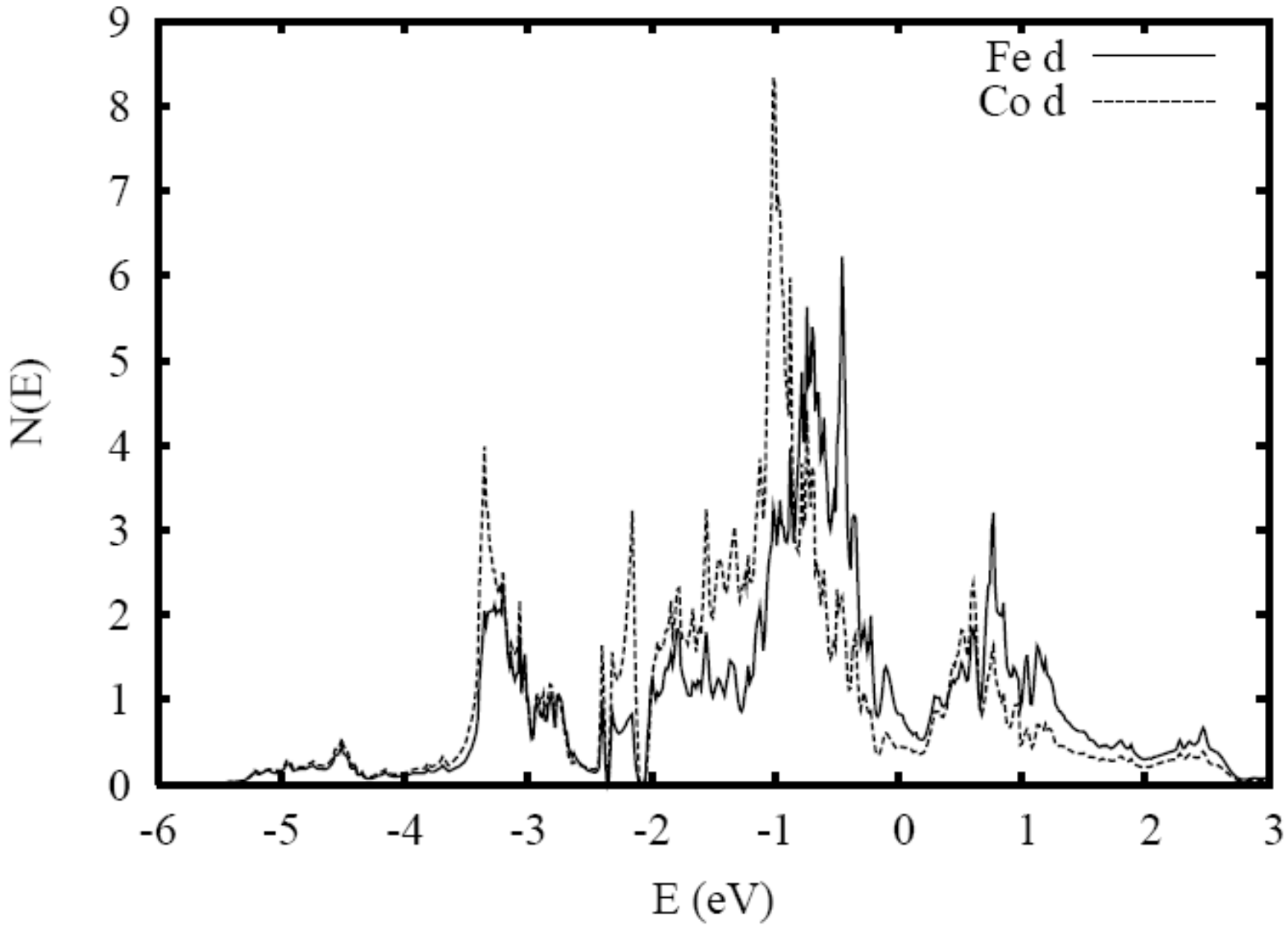

(b)

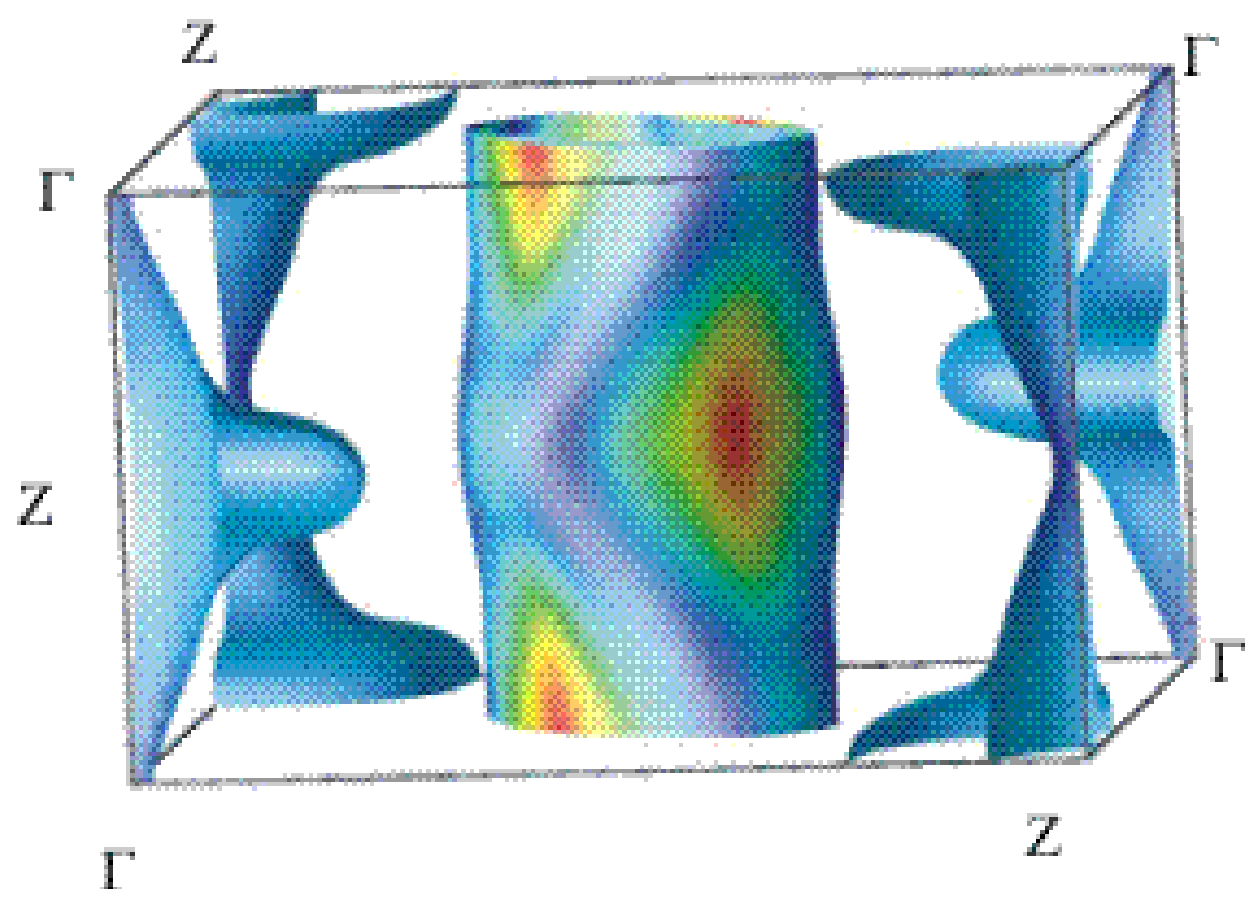

Fig. 3. (Color online) (a) Projections of the *d* electron DOS, on a per atom basis, for the Fe and Co in a $Ba_4Fe_7CoAs_8$ supercell. (b) LDA Fermi Surface of $BaFe_{1.8}Co_{0.2}As_2$, for the LDA internal coordinates, shaded by band velocity (light blue is the low velocity).